\documentclass[aps,prl,twocolumn,groupedaddress,showpacs]{revtex4}

\usepackage{graphicx}

\begin{document}



\title{Frequency Dependence and Dissipation in the Dynamics of Solid Helium}


\author{Oleksandr Syshchenko}
\author{James Day}
\author{John Beamish}
\affiliation{Department of Physics, University of Alberta,
Edmonton, Alberta, Canada, T6G 2G7}


\date{\today}

\begin{abstract}
Torsional oscillator experiments on solid $^4$He show frequency changes which suggest mass
decoupling, but the onset is broad and is accompanied by a dissipation peak.
We have measured the elastic shear modulus over a broad frequency range,
from 0.5 Hz to 8 kHz, and observe similar behavior - stiffening and a dissipation peak.
These features are associated with a dynamical crossover in a
thermally activated relaxation process in a disordered system rather than a true phase
transition.  If there is a transition in the DC response, e.g. to a supersolid state, it
must occur below 55 mK.
\end{abstract}


\maketitle



As the most quantum of solids, helium is a likely candidate for supersolidity but it was not until 2004 that experimental evidence for supersolidity appeared in torsional oscillator (TO) measurements\cite{Kim04-1941, Balibar08-173201}.  The key observation was an increase in the TO frequency at temperatures below 200 mK, suggesting that some of the solid helium decoupled from the oscillator, the  \textquotedblleft non-classical rotational inertia" (NCRI)  which characterizes a supersolid.  However, the onset of NCRI in TO experiments is  broad and is always accompanied by a dissipation peak, features typical of a dynamical crossover in a relaxation process.  Does the gradual onset represent the DC response associated with a phase transition to a new state, broadened perhaps by disorder, or is it a frequency dependent effect, for example due to flow in a vortex liquid\cite{PWAnderson07-160} or glass-like relaxation\cite{Nussinov07-014530, Graf10, Hunt09-632}?  Measurements over a broad frequency range can distinguish between these possibilities and help identify possible relaxation mechanisms, but are almost impossible with TOs.

We have studied the mechanical susceptibility of solid helium directly, by measuring the real and imaginary parts of its elastic response (shear modulus and dissipation), analogous to the TO frequency and dissipation.  This has advantages over a TO: the measurement is direct, the modulus changes are large and, most importantly, the technique is non-resonant so we could measure $\mu$ over four orders of magnitude in frequency. We see a dissipation peak associated with the previously reported\cite{Day07-853, Day09-214524} shear stiffening.  The peak and the onset of stiffening shift to lower temperatures as the frequency decreases, with an activation energy around 0.7 K.  A broad range of activation energies is needed to fit the data - the relaxation must occur in a highly disordered system.

The frequency of a TO is usually considered to be a direct probe of moment of inertia (and thus of any mass decoupling) but things are more complicated when the helium is solid.  Helium affects a TO through the \textquotedblleft back action" force it exerts on its walls\cite{Nussinov07-014530}, which could be purely elastic or might require a more complicated model\cite{PWAnderson07-160, Nussinov07-014530, Yoo09-100504, Hunt09-632,Graf10}.  The period and damping of the TO (real and imaginary parts of the angular susceptibility) are determined by the magnitude and phase of this force.  A decrease in the helium's moment of inertia reduces the elastic stress at the walls and increases the frequency.  An increase in its shear modulus would also raise the TO frequency by making the composite TO/helium system stiffer, thus mimicking mass decoupling, but this effect is too small to explain the observed frequency changes\cite{Clark08-184531}.  Elastic forces are in phase with the TO drive and affect only the real part of the response, but any out of phase stresses will produce dissipation.  Dissipation, e.g. the peak which accompanies NCRI\cite{Kim04-1941, Rittner06-165301, Aoki07-015301, Hunt09-632}, will affect the TO frequency even without mass decoupling\cite{Nussinov07-014530}.  However, the frequency change predicted for a simple Debye relaxation peak is nearly always much smaller than the observed increase, suggesting that most of the apparent decoupling may still be due to supersolidity\cite{Hunt09-632}.  More complicated response functions give better fits to the TO data\cite{Graf10}, but involve additional parameters.  Varying the measurement frequency would provide a much more stringent test of any model of the helium's response but this is difficult with a TO.  It has been possible to operate a TO off resonance, but the frequency range was very limited (575 Hz +/- 10$\%$)\cite{Davis09-private}. The clearest evidence of frequency dependence comes from a TO designed with two modes\cite{Aoki07-015301} - at 496 Hz, the NCRI began at lower temperatures than at 1173 Hz.

Elastic measurements\cite{Day07-853, Day09-214524, Sasaki07-205302, Mukharsky09-140504} on solid helium show effects which are clearly related to the TO behavior.  The shear modulus $\mu$ increases by as much as 20$\%$, with the same dependence on temperature, amplitude and $^3$He impurity concentration as the NCRI\cite{Day07-853}.  This confirms the unusual nature of solid $^4$He, although these measurements do not directly address the question of whether the low temperature state is supersolid.  In this paper we extend our shear modulus measurements\cite{Day07-853, Day09-214524} to much lower frequencies, below 1 Hz, and we simultaneously measure the dissipation.  In some cases, we also measured the frequency $f_{r}$ and width of an acoustic resonance, extending our frequency range to 8 kHz.

Measurements were made as in ref. 8.  Crystals were grown from $^4$He (0.3 ppm $^3$He) over about 1 hour (using the blocked capillary technique), filling a gap of thickness D (0.2 to 0.5 mm) between two transducers with area A (1 cm$^2$).  An AC voltage $V$ with angular frequency $\omega = 2 \pi f$, was applied to one transducer (with piezoelectric coefficient $d_{15}$), generating a shear displacement  $\delta x = d_{15}V$ at its surface, a strain $\epsilon =\delta x/D$ in the helium and a stress $\sigma$ on the detecting transducer.  This produced a charge $q$ which was measured as a current $I = \omega q$, giving the shear modulus as  $\mu=\frac{\sigma}{\epsilon}=(\frac{D}{2\pi d_{15}^{2}A})(\frac{I}{fV})$.  For purely elastic deformations, stress is in phase with the applied strain but in general the stress lags the strain by a phase angle $\phi$ which is related to the dissipation by $1/Q =$ tan $\phi$ ($\approx \phi$ for small dissipation).  The amplitude and phase of the current thus give the real and imaginary parts of the helium's shear modulus, although electronic phase shifts prevent us from measuring absolute values of 1/Q.  All data in this paper were taken in the linear response regime, at strains $\epsilon < 10^{-8}$ where there is no amplitude dependence or hysteresis\cite{Day07-853, Day10}.  We measured the modulus at frequencies between 0.5 and 2500 Hz. 				

Figure 1a compares the shear modulus $\mu$ and acoustic resonance frequency $f_r$ (for a 33.3 bar sample) with the torsional oscillator NCRI (for a 65 bar sample\cite{Kim04-1941}).  Figure 1b shows the corresponding dissipation, determined from the phase angle between stress and strain, the width of the resonance peak\cite{Day09-214524} and the TO amplitude.   The shear modulus (measured at 200 Hz) and the NCRI (at 910 Hz) both increase below 200 mK, with very similar temperature dependence, and both have dissipation peaks below 100 mK.  The acoustic resonance shows essentially the same behavior.  The increase in $f_r$ is half as large as the increase in $\mu$ (3.7$\%$ vs. 7.4$\%$), as expected since $f_r$ scales with the transverse sound speed, i.e. as $\sqrt{\mu}$.   The acoustic resonance dissipation peak is roughly the same size as for the modulus but is broader and shifted to higher temperature.
\begin{figure}
\includegraphics[width=\linewidth]{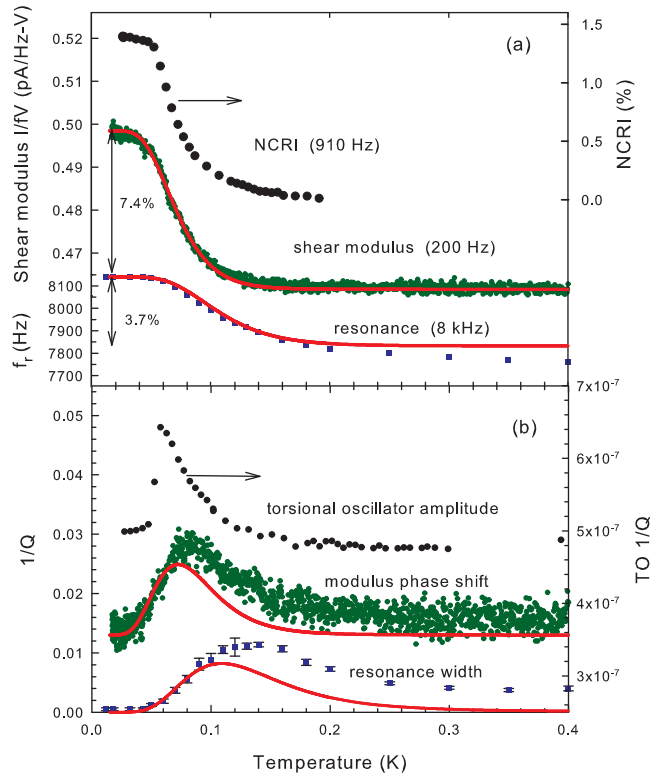}
\caption{Mechanical response in the 33.3 bar (20.4 cm$^3$/mole)$^4$He crystal from ref. 7. a) Shear modulus (at 200 Hz) and acoustic resonance frequency (left axis) and TO NCRI (right axis).  b) Dissipation corresponding to Fig. 1a. Solid (red) lines are fits described later in the text. The 200 Hz dissipation and fit are vertically offset for clarity.}
\label{fig:Figure1.EPS}
\end{figure}
Figure 2 shows the frequency dependence of $\mu$ and 1/Q in a 38 bar crystal.  Similar behavior was observed in other samples\cite{Day07-853}, including the crystal of Fig. 1.  The total modulus change is independent of frequency.  It is larger, $\frac{\delta\mu}{\mu_o}\sim$15$\%$, than that for the crystal of Fig. 1 but the temperature dependence is similar.  A dissipation peak is centered near the temperature where the modulus is changing most rapidly.  The peak and the modulus change shift to lower temperatures as the frequency decreases - the behavior expected for a thermally activated relaxation process.  For a simple Debye process with relaxation time $\tau$, the modulus $\mu$ and dissipation $1/Q$ are related to the real and imaginary parts of the shear modulus\cite{Nowick72}
\small
\begin{equation}
    \frac{\mu}{\mu_{o}}=1-\frac{\delta\mu}{\mu_{o}}\frac{1}{1+(\omega\tau)^{2}}
\end{equation}

\begin{equation}
    \frac{1}{Q}=\frac{\delta\mu}{\mu_{o}}\frac{\omega\tau}{1+(\omega\tau)^{2}}	   	
\end{equation}
\normalsize
where $\mu_o$ is the \textquotedblleft unrelaxed modulus" ($\omega\tau\gg$1) and $\mu_o - \delta\mu$ is the \textquotedblleft relaxed modulus" ($\omega\tau\ll$1).  The strength of the relaxation determines the total change $\frac{\delta\mu}{\mu_o}$.  For a thermally activated process, $\tau (E) = \tau_{o} e^{\frac{E}{T}}$ where E is the activation energy.  The crossover from unrelaxed to relaxed modulus occurs at the temperature where $\omega\tau=1$; at this point the dissipation is maximum and 50$\%$ of the modulus change has occurred.  These points (indicated by circles in Fig. 2) can be used to determine the temperature dependence of $\tau$.
\begin{figure}
\includegraphics[width=\linewidth]{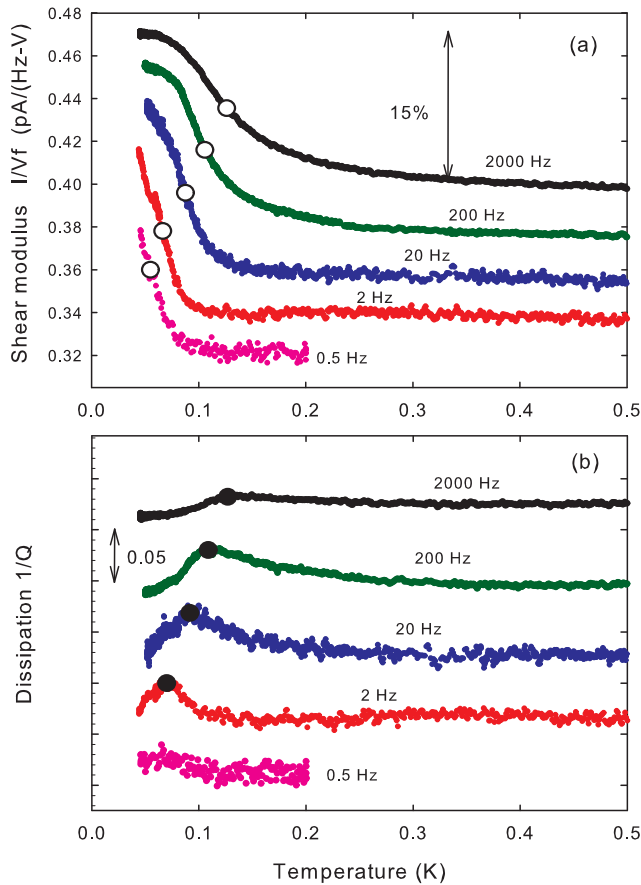}
\caption{Frequency dependence of a) shear modulus and b)  dissipation for an hcp $^4$He crystal at 38 bar (20.1 cm$^3$/mole). Curves have been vertically shifted for clarity (the modulus scale is for the 2000 Hz data). Dots are transition midpoints plotted in Fig. 3.}
\label{fig:Figure2.EPS}
\end{figure}
Figure 3 is an Arrhenius plot (1/T vs log $\omega$) of the crossover temperatures marked on Fig. 2.  We also show the corresponding points (plus a point from the acoustic resonance dissipation peak at 8 kHz) for the 33.3 bar crystal of Fig. 1.  Although there is some scatter, the relaxation processes are clearly thermally activated.  The data for the 38 and 33.3 bar samples have similar slopes, corresponding to activation energies  $E \sim$ 0.77 K and 0.73 K, respectively, but are offset vertically, indicating that they have different attempt times $\tau_{o}$ (consistent with sample to sample variations\cite{Day07-853})
\begin{figure}
\includegraphics[width=\linewidth]{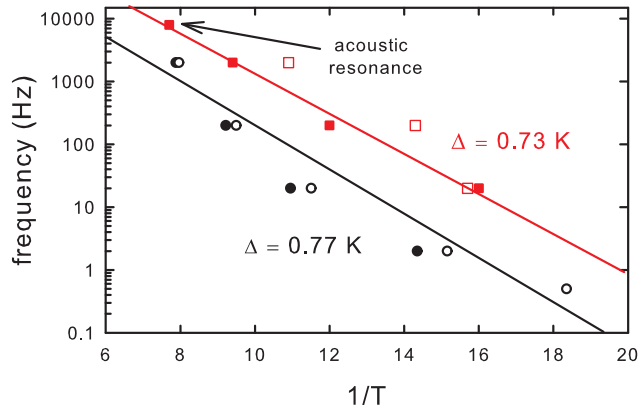}
\caption{Frequency dependence and thermal activation for the crystals of Figs. 1 and 2.  Open symbols are the temperatures at which 50$\%$ of the modulus change has occurred, closed symbols are for the dissipation peaks (as in Fig. 2).  Lines correspond to activation energies of 0.73 and 0.77 K.}
\label{fig:Figure3.EPS}
\end{figure}
We can check whether a simple Debye relaxation describes the modulus and dissipation in solid $^4$He by comparing eqns. (1) and (2) to the 200 Hz data of Fig. 1.  The dashed (blue) line in Fig. 4a is the modulus at 200 Hz, calculated from eqn. (1) using the value of the activation energy from Fig. 3 (E=0.73 K) and the modulus change $\frac{\delta\mu}{\mu_o}=0.074$.  The value of $\tau{_o}$ (25 ns) is chosen to make the modulus change midpoint agree with the data.  It is clear that the actual crossover is much broader than for a simple Debye relaxation.  We can get a better fit to the modulus data with a smaller activation energy (requiring a larger attempt time). The dotted (black) curve in Fig. 4a is for E=0.175 K and $\tau{_o}$=65 $\mu$s.  Despite the good fit, it cannot be correct.  First, this value of E is inconsistent with the frequency dependence shown in Fig. 3.  Second, the predicted dissipation peak (dotted black curve in Fig. 4b) is too large by a factor of 2.5.  This is also true for the fit with E=0.73 K and $\tau{_o}$=25 ns (dashed blue curve), which in addition is much narrower than the measured peak.  No choice of E and $\tau{_o}$ can resolve this discrepancy since eqns. (1) and (2) imply that the total modulus change and the height of the dissipation peak are directly related by $(\frac{1}{Q})_{peak}=\frac{1}{2}\frac{\delta\mu}{\mu_{o}}$.

However, if the relaxation process has a distribution of activation energies rather than a single value, then the crossover will be broader and the height of the dissipation peak will be reduced (ref. 17, Chapter 4).  We can fit the 200 Hz modulus data using a distribution of activation energies with characteristic energy $\Delta$ and width $W$
\small
\begin{equation}
    n(E)= B e^{-{\frac{(ln E -ln \Delta)^{2}}{W^{2}}}}
\end{equation}
\normalsize
where B is the normalization factor.  This distribution has a long tail at large E and approaches zero for small E.  The modulus and dissipation are found by multiplying n(E) by the contributions from eqns. (1) and (2) and integrating over all energies E.

The solid (red) curve in Fig. 4a shows the modulus using this distribution with $\Delta=0.73 K$ and $W$=0.45.  The fit is much better than for a single activation energy of 0.73 K (dashed blue curve).  The predicted dissipation peak (solid red curve in Fig. 4b) has approximately the right height and width, although there is additional dissipation above the peak (nearly constant above 0.25 K).  This distribution also gives the correct frequency dependence. The solid (red) curves in Fig. 1 show the predicted frequency $f_{r}$ and dissipation for the acoustic resonance at 8 kHz (and for $\mu$ and 1/Q at 200 Hz) using these parameters ($\Delta=0.73 K$, $\tau_{o}$=9 ns and $W$=0.45).  The fit to $f_{r}$ is good, showing that frequency dependence retains its Arrhenius form with the broadened distribution of activation energies.  The predicted dissipation peak for the acoustic resonance has approximately the correct height and width but again there is additional dissipation at temperatures above the peak.
\begin{figure}
\includegraphics[width=\linewidth]{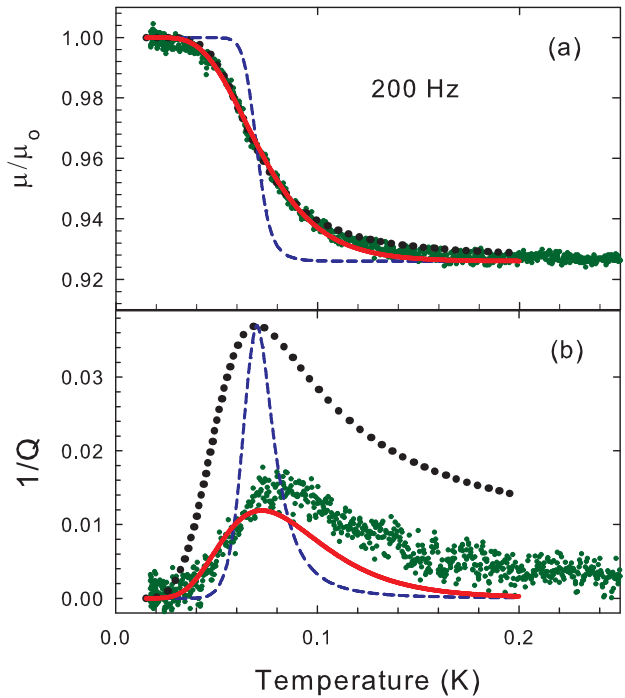}
\caption{Fits of thermally activated relaxation process to a) the shear modulus and b) the dissipation at 200 Hz for the 33.1 bar crystal of Fig. 1.  Fits are explained in the text.}
\label{fig:Figure4.EPS}
\end{figure}

In both TO and shear modulus measurements the change in the real part of the response is accompanied by damping and the onset temperature decreases as the frequency is reduced\cite{Aoki07-015301}, but the dissipation peak is significantly smaller than a simple Debye relaxation would predict\cite{Hunt09-632}.  This behavior is consistent with a thermally activated process if there is a broad range of relaxation times\cite{Graf10}.  In contrast to TO measurements, the wide frequency range of our measurements allows us to reliably determine the characteristic activation energy for the first time and, using this value, $\Delta \sim 0.7 K$, to fit the temperature and frequency dependence of both the modulus and dissipation using a distribution of activation energies.   The onset of stiffening and the broad dissipation peak are not signatures of a phase transition near 150 mK, but rather of a frequency dependent crossover in an activated relaxation process.  The temperature of this crossover continues to decrease with frequency down to at least 55 mK (at 0.5 Hz) - if there is a true phase transition reflected in, e.g., the DC elastic response, it must be at a lower temperature.

The effects of annealing on shear modulus and TO behavior\cite{Day09-214524, Rittner06-165301} show that defects are involved.  It is also clear that $^3$He impurities are important since the onset of stiffening\cite{Day07-853} (and of TO decoupling\cite{Kim08-065301}) shifts to lower temperature with reduced $^3$He concentration (in recent experiments with extremely low $^3$He concentrations\cite{Rojas10} the onset of stiffening was at even lower temperature, close to 40 mK).  The large shear modulus changes we observe are very difficult to understand except from dislocation motion\cite{Day07-853, Day09-214524, Nowick72} and the role of $^3$He atoms is to bind to and immobilize dislocations at low temperatures.  As the temperature is raised, they unbind thermally, allowing the dislocations to move and relax the applied stress, thus reducing the shear modulus.  In this picture, the observed activation energy is the binding energy for a $^3$He atom on a dislocation; the value $\Delta \sim 0.7 K$ is consistent with estimates from previous experiments\cite{Paalanen81-664, Iwasa80-1722}.  The value of $\tau_{o}$=9 ns, is comparable to the inverse of a thermal frequency, $k_{B}T/h$ (1 ns at 50 mK). The broad range of energies needed to fit our data indicates that the relaxation is occurring in highly disordered network of pinned dislocations. This system might exhibit glassy dynamics, as has been suggested based on the TO response\cite{Graf10}.  The "blocked annulus" experiment\cite{Kim04-1941}, which implies that the decoupling seen in TO experiments is associated with long range order, cannot be explained by local dynamics of dislocation motion and remains the strongest evidence for supersolidity.  Our measurements cannot rule out an intrinsic phase transition, e.g. one associated with roughening due to kinks on dislocations\cite{Aleinikava08}, but it would have to occur below 55 mK.

This work was supported by NSERC Canada.

\bibliography{supersolid}

\end{document}